\documentclass{emulateapj}
\usepackage{graphicx}
\usepackage{epstopdf}
\usepackage{color}

\newcommand{\Mjup}{\mbox{$M_\mathrm{Jup}$}}
\newcommand{\Msun}{\mbox{$M_{\odot}$}}

\begin{document}
\title{Planets Around Low-Mass Stars (PALMS). \\ II.  A Low-Mass Companion to the Young M Dwarf GJ~3629 Separated By 0$\farcs$2*}
\author{Brendan P. Bowler,\altaffilmark{1} 
Michael C. Liu,\altaffilmark{1} 
Evgenya L. Shkolnik,\altaffilmark{2}
and Motohide Tamura\altaffilmark{3}
\\ }
\email{bpbowler@ifa.hawaii.edu}

\altaffiltext{1}{Institute for Astronomy, University of Hawai`i; 2680 Woodlawn Drive, Honolulu, HI 96822, USA}
\altaffiltext{2}{Lowell Observatory, 1400 W. Mars Hill Road, Flagstaff, AZ 86001}
\altaffiltext{3}{National Astronomical Observatory of Japan, 2-21-1 Osawa, Mitaka, Tokyo 181-8588, Japan}
\altaffiltext{*}{Some of the data presented herein were obtained at the W.M. Keck Observatory, which is operated as a scientific partnership among the California Institute of Technology, the University of California and the National Aeronautics and Space Administration. The Observatory was made possible by the generous financial support of the W.M. Keck Foundation.}

\submitted{ApJ, in press}
\begin{abstract}

We present the discovery of a 0$\farcs$2 companion to the young M dwarf GJ 3629 as part
of our high contrast adaptive optics imaging search for giant planets around low-mass stars with the Keck-II 
and Subaru telescopes.  Two epochs of imaging confirm the pair is co-moving and reveal signs of 
orbital motion.  The primary exhibits saturated X-ray emission, which together with its UV photometry 
from $GALEX$ point to an age younger than $\sim$300~Myr.  
At these ages the companion lies below the hydrogen
burning limit with a model-dependent mass of 46~$\pm$~16~\Mjup \ based on the system's photometric  
distance of 22~$\pm$~3~pc.  Resolved $YJHK$ photometry 
of the pair indicates a spectral type of M7~$\pm$~2 for GJ~3629~B.  
With a projected separation of 4.4~$\pm$~0.6~AU and an estimated orbital 
period of 21~$\pm$~5~yr, GJ 3629~AB is likely to yield a dynamical mass in the next several years, making it one of 
only a handful of brown dwarfs to have a measured
mass and an age constrained from the stellar primary.

\end{abstract}

\section{Introduction}{\label{sec:intro}}

Following a decade of attempts to directly image extrasolar giant planets with increasingly
sensitive instruments and speckle suppression techniques (e.g., \citealt{Lowrance:2005p18287}; 
\citealt{Biller:2007p19401}; \citealt{Lafreniere:2007p17991}; \citealt{Liu:2010p21647}), several planetary systems
have finally been found through high contrast imaging over the past few years.  These discoveries
have opened up an exciting new era of planetary science where
the atmospheres of non-transiting extrasolar planets can be directly studied for the first time.  
Five unambigious planets have been imaged orbiting two stars: one around the A5 star
$\beta$~Pic (\citealt{Lagrange:2009p14794}; \citealt{Lagrange:2010p21645}), 
and four surrounding the A5 star HR~8799 (\citealt{Marois:2008p17990}; 
\citealt{Marois:2010p21591}).\footnote{Because the optically-detected 
companion to Fomalhaut (\citealt{Kalas:2008p18842}) has not been recovered 
in the infrared (\citealt{Marengo:2009p18897}), it is unclear 
whether the object is a planet, perhaps with a large high-albedo ring system, or something else, like a dust cloud
from a recent planetesimal collision (\citealt{Janson:2012p23488}).  
Recently \citet{Kraus:2012p23492} have discovered what appears to be a 
young ($\sim$2~Myr) accreting giant planet orbiting the transition-disk star 
LkCa~15 at $\sim$15--20~AU, but a clear understanding
of this system, including the mass of the companion, is still lacking.
 In addition to these objects, a handful of other 
planetary-mass companions have been found orbiting stars from hundreds 
(e.g., 1RXS~J1609--2105~b: \citealt{Lafreniere:2008p14057}; 
GSC~06214--00210~b: \citealt{Ireland:2011p21592}) 
to thousands (Ross 458~C: \citealt{Goldman:2010p22044}, \citealt{Scholz:2010p20993}; 
WD~0806-661~B: \citealt{Luhman:2011p22766}) 
of AU, but the formation mechanism of these enigmatic companions 
remains obscure (e.g., \citealt{Bowler:2011p23014}).}
The fact that all these planets orbit high-mass stars might at first suggest that giant planet
formation is more efficient around massive stars, which is a well-established trend observed
at smaller separations from radial velocity planet searches (\citealt{Johnson:2007p169};
\citealt{Lovis:2007p17712}; \citealt{Bowler:2010p19983}; \citealt{Johnson:2010p20950}).  
However, high contrast imaging
searches have mostly neglected low-mass stars.   Until recently there has been
a scarcity of known nearby young M dwarfs, making it difficult to produce statistical comparisons
of planet occurrence rates as a function of stellar mass.  As a result, even though M dwarfs 
outnumber AFGK stars by a factor of 2--3 in the solar neighborhood (\citealt{Henry:1997p22864}; \citealt{Bochanski:2010p23010}), 
our understanding of planet formation is the weakest in this stellar mass regime.

We are conducting a high contrast adaptive optics 
imaging survey of young M dwarfs with Keck-II and Subaru to search for young giant planets and brown dwarfs 
and measure the frequency of giant planets orbiting low-mass stars.  
Our targets are newly identified M dwarfs with ages $<$300~Myr
and distances $\lesssim$30~pc which were selected based on elevated levels of X-ray and UV emission
(\citealt{Shkolnik:2009p19565}; \citealt{Shkolnik:2011p21923}).  
Compared to high-mass stars, M dwarfs present several advantages as targets for
direct imaging.
Their higher space densities mean they are on average closer than
more massive stars, so smaller physical separations can be probed.  
Moreover, because M dwarfs are intrinsically faint, direct imaging can detect lower planet masses 
in the contrast-limited regime.  This allows us to reach
typical planet masses of a few \Mjup \ at separations of $\sim$10~AU, making 
our Planets Around Low-Mass Stars (PALMS) survey one
of the deepest direct imaging planet searches to date.

The first discovery from our PALMS survey was an L0 substellar companion 
to the young M~dwarf 1RXS~J235133.3+312720 separated by $\sim$120~AU
(\citealt{Bowler:2012p23698}).  Here we present the discovery of a 0$\farcs$2 companion to the 
young M3.0 star GJ~3629. 
\citet{Shkolnik:2009p19565} identified GJ~3629 as a young star based
on its high fractional X-ray luminosity, which is comparable to members of the Pleiades ($\sim$125~Myr)
and young moving groups (10--100~Myr).  
 \citet{Shkolnik:2009p19565} obtained a high resolution optical spectrum of GJ~3629~A
and did not detect Li, thereby setting a lower limit to its age.  Based on these constraints they estimate
an age of 25--300~Myr, which places our new companion GJ~3629~B below the 
hydrogen burning minimum mass at the estimated distance of the system (22~$\pm$~3~pc; Section~\ref{sec:dist}).

\section{Keck-2/NIRC2 NGS AO Imaging}{\label{sec:obsnirc2}}

\begin{figure}
  \begin{center}
  \resizebox{5in}{!}{\includegraphics{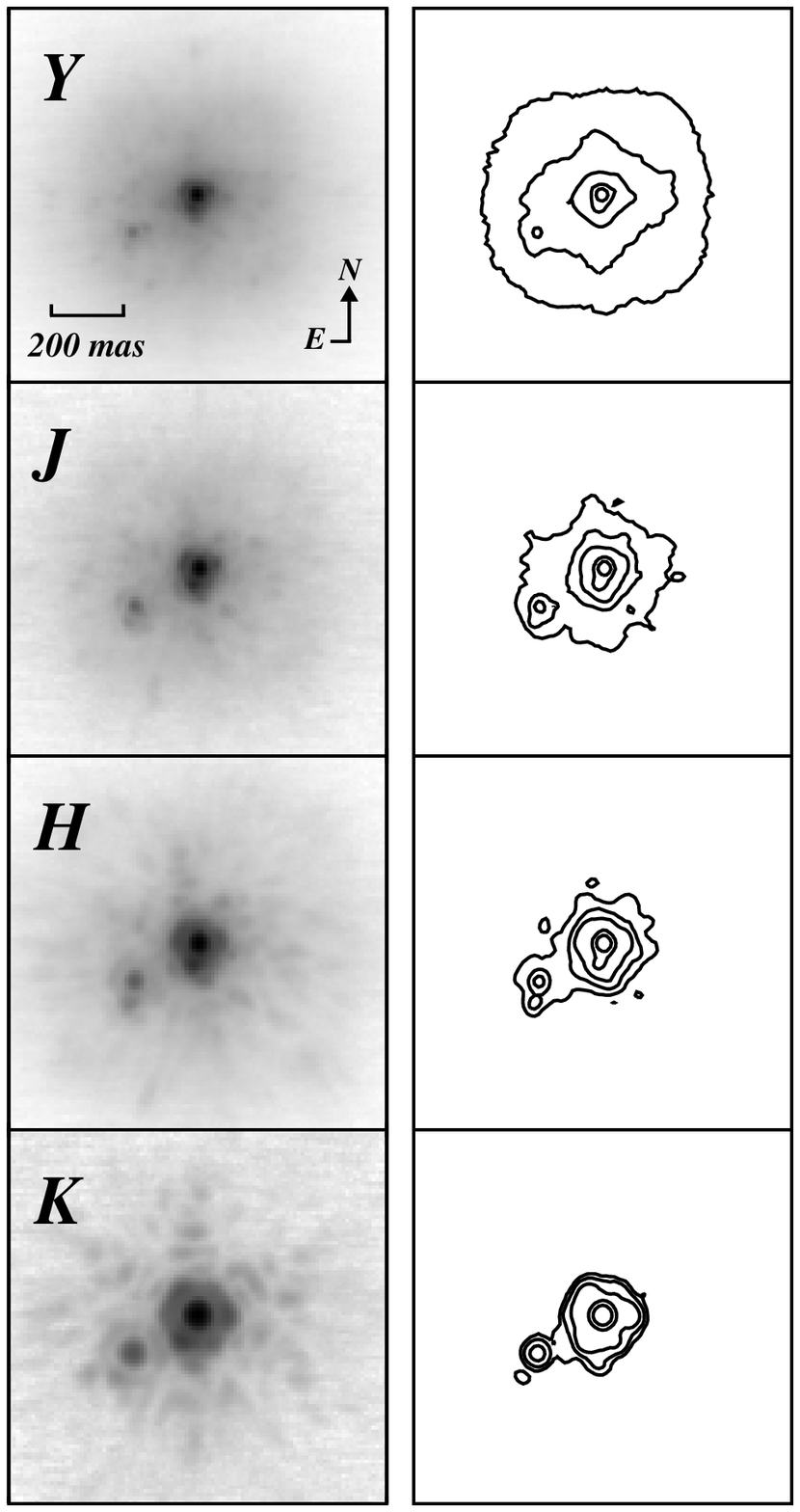}}
  \caption{Keck-II/NIRC2 $YJHK$ images of GJ~3629~AB.  The left panels show
  coadded frames of the system with an asinh stretch to bring out faint features (\citealt{Lupton:2004p20516}).  
  The right panels show contours
  representing 50\%, 20\%, 5\%, 2\%, and 1\% of the peak flux from the primary.  Artifacts from the AO correction
  during the poor second epoch conditions are visible below the real point sources, 
  especially in the $J$- and $H$-band frames.  North
  is up and East is left in the images.   \label{fig:contours} } 
\end{center}
\end{figure}

We imaged GJ~3629 on 25 March 2011 UT with the Near Infrared Camera~2 (NIRC2) in its narrow-field mode on Keck-II
coupled with natural guide star adaptive optics (NGS~AO; \citealt{Wizinowich:2000p21634}).  
There were light cirrus clouds during the observations and the seeing 
was $\sim$0$\farcs$6 as reported by the Differential Imaging Motion Monitor on CFHT.  
We obtained three short dithered frames with the  $K_S$ filter while reading out the 
central 512$\times$512 pixels.  GJ~3629 was easily resolved into a $\sim$200~mas pair 
with a flux ratio of $\sim$3~mag (the diffraction limit in $K$ is $\sim$55~mas).
Second-epoch observations were obtained on 3 March 2012 UT with Keck-II/NIRC2 (narrow-field) and NGS~AO.
Humidity levels were near 100\% most of night and there were several bouts of snow,
but for about half an hour the humidity dropped and the sky was clear enough to observe 
GJ~3629 in order to verify the companion was comoving with the primary.  
We imaged the system in dithered patterns with the $YJHK$ filters (Figure~\ref{fig:contours}),
which are from the Mauna Kea Observatory filter system 
(\citealt{Simons:2002p20490}; \citealt{Tokunaga:2005p18542}).
A summary of our observations are listed in Table~\ref{tab:obs}.

The images were reduced in the standard fashion by removing bad pixels, subtracting 
dark frames, and dividing by normalized dome flats obtained 
at the start of the night.  
Each image was North-aligned using the FITS header keywords, taking into account the 
AO-detector offset (+0.7$^{\circ}$) and the sky orientation on the detector (+0.252$^{\circ}$) 
derived by \citet{Yelda:2010p21662}.
We computed astrometry and flux ratios by fitting an analytic model composed of
three elliptical Gaussians to each binary component as described in \citet{Liu:2008p14548}.
We found that varying the number of Gaussians in the input PSF model (two versus three) 
resulted in systematic errors in the flux ratios of $\approx$1\%, so we incorporated this in our flux ratio uncertainty 
by adding it in quadrature with the measured random errors. 
No distortion correction was applied because the relative correction over $\sim$20~pixels is 
negligible for our purposes.
For the separation measurement we adopt the NIRC2 plate scale of 9.952~$\pm$~0.002 mas pix$^{-1}$ derived by  \citet{Yelda:2010p21662}.
The resulting astrometry and flux ratios are listed in Table~\ref{tab:astrometry}, where the quoted values and uncertainties 
represent the means and standard deviations of each data set.  Strehl ratios and full width at half-maximum (FWHM)
measurements of GJ~3629~A were made with the publicly available IDL routine \texttt{NIRC2STREHL}
and are listed alongside the astrometry in Table~\ref{tab:astrometry}.

\section{Results}{\label{sec:results}}

\subsection{Distance}{\label{sec:dist}}

There is no parallax measurement for GJ~3629~A but there are several distance estimates to the primary in the literature.
GJ~3629 is part of the $NStars$ sample of nearby stars ($\lesssim$20~pc; \citealt{Reid:1995p22125}; \citealt{Reid:2004p20383}) 
and was assigned a spectrophotometric distance of 16.7~$\pm$2.7~pc (\citealt{Shkolnik:2009p19565}) 
based on optical bandstrength-absolute magnitude relations.  
Recently Shkolnik et al. (2012, submitted) measured 
parallaxes for 75 young ($<$300~Myr) M-type stars with spectrophotometric distance estimates $<$30~pc.
They found that the photometric distances were typically underestimated by $\sim$70\%
as a result of unresolved binarity and youthful overluminosity, which would place GJ~3629~AB at $\sim$28~pc.  
Shkolnik et al. (2012, submitted) also derive a distance estimate using the minimum (25~Myr)
and maximum (300~Myr) age estimates of the system (Section~\ref{sec:age}) combined with the 
evolutionary models of \citet{Baraffe:1998p160},
arriving at a value of 22~$\pm$~7~pc.
Other independent distance estimates for the system include photometric distances of 18$^{+8}_{-4}$~pc and 
22.2~$\pm$~5.6~pc by \citet{Lepine:2011p22863} and \citet{Lepine:2005p70}, 
which were computed using color-absolute magnitude relations from field M dwarfs, and an estimate of 
26~pc by \citet{Wright:2011p23132} using 1~Gyr theoretical isochrones from \citet{Siess:2000p11017}.
Given this large range of distance estimates, we adopt a value of 22~$\pm$~3~pc for GJ~3629~AB.

\subsection{Common Proper Motion}

\begin{figure}
  \resizebox{\textwidth}{!}{\includegraphics{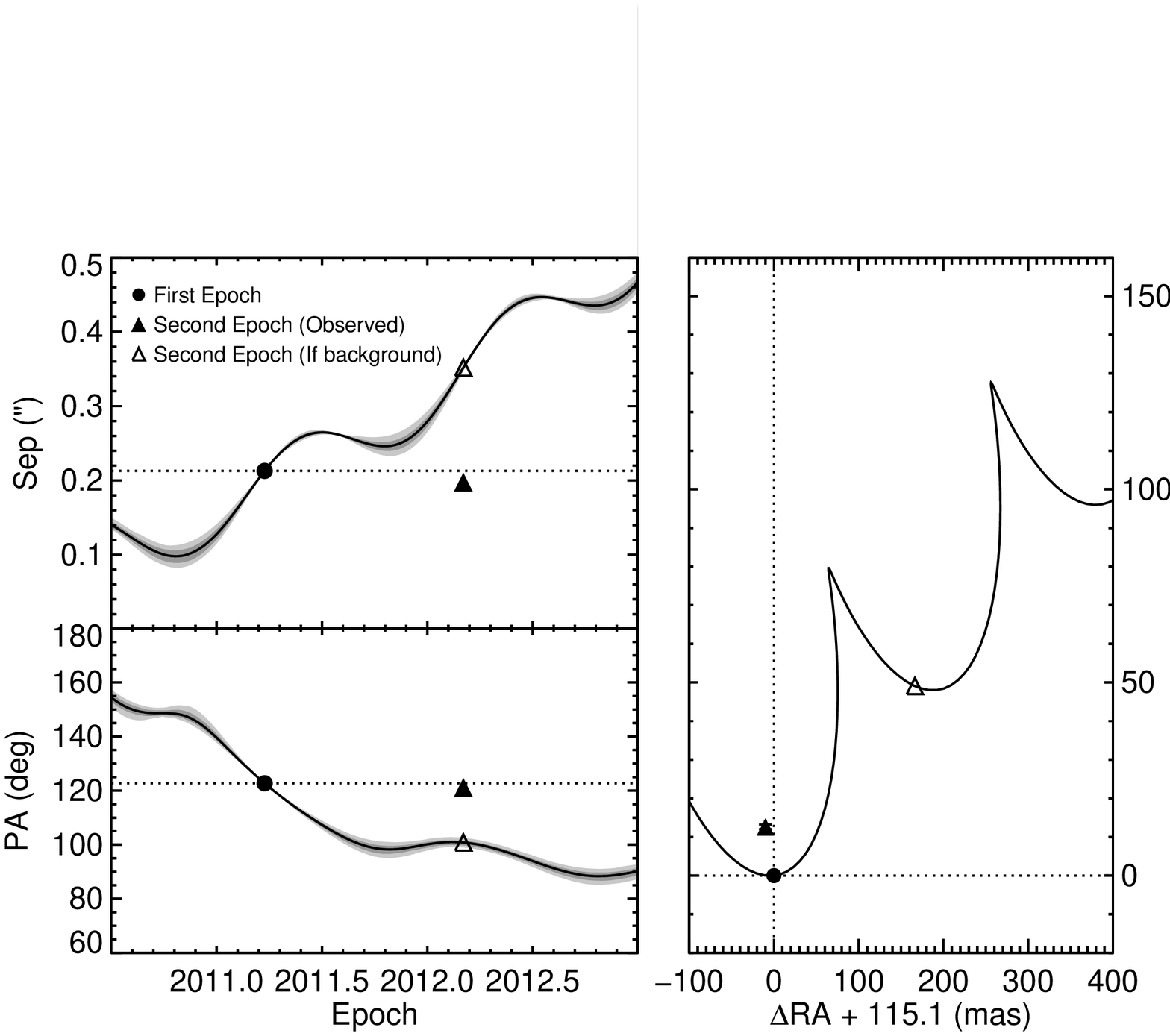}}
  \caption{Predicted relative motion of a stationary object at the first epoch 
  position (filled circle) of the companion to GJ~3629~A.  The left panels shows the change 
  in separation and position angle 
  over time due to the parallactic and proper motion of the primary.  Gray 1- and 2-$\sigma$ shaded 
  confidence intervals incorporate uncertainties in the proper motion, distance estimate (22~$\pm$~3~pc),
  and first epoch astrometry. The right panel
  shows the motion in relative RA and Dec offsets ($\Delta$ refers to primary -- secondary position).
  Our second epoch measurement (filled triangle) rules out the background hypothesis (open triangle)
   and shows the companion shares nearly
  common proper motion with the primary, with the slight difference between epochs attributable to
   orbital motion.  The astrometric measurement uncertainties are smaller than the size of the symbols.  \label{fig:tracks} } 
\end{figure}

The baseline of 11 months between our first and second epoch observations 
lets us easily distinguish whether the close companion is 
physically bound to the primary.   
Figure~\ref{fig:tracks} shows the expected relative motion
of a stationary background object based on the proper and parallactic motion of GJ~3629.
The gray 1- and 2-$\sigma$
confidence intervals for the background tracks in Figure~\ref{fig:tracks} are
based on Monte Carlo realizations incorporating
uncertainties in the 
distance estimate, proper motion, and astrometry.
The weighted mean of the second epoch astrometry from our four independent
measurements (one for each filter) is 197.9~$\pm$~0.2~mas and 
120.75~$\pm$~0.07$^{\circ}$.
If the candidate companion was a distant background object then the expected separation
(352~$\pm$~3~mas) and position angle (100.8~$\pm$~0.9$^{\circ}$) 
at the time of our second epoch observation
would differ from the measured values by 51-$\sigma$ and 22-$\sigma$, respectively, 
easily ruling out the background hypothesis.\
Instead, our second epoch astrometry is much closer to the first epoch measurements,
differing by 15.0~$\pm$0.2~mas and 1.96~$\pm$~0.11$^{\circ}$.  This difference is plausibly
explained by orbital motion.  

While this rules out contamination from a stationary background object, there is a small (but non-zero)
probability that the candidate companion is an unassociated star with 
coincident sky position and proper motion.  We approach this problem in two ways:
(1) by calculating the \textit{a priori} probability of a chance alignment of a field ultracool dwarf
regardless of any information about proper motions, and (2) using a galaxy model to
derive the expected contamination rate of background stars with proper motions similar to GJ~3629.

The probability of an ultracool dwarf falling in the NIRC2 field of view can be computed from the surface density of ultracool dwarfs
($\Sigma_{UC}$) and the solid angle subtended by the instrument ($\Omega$).  The most complete census of ultracool dwarfs within
20 pc was compiled by \citet{Reid:2008p20073}.  They identified 196 M7--T2.5 systems covering $\approx$65\% of
the celestial sphere, which corresponds to a surface density $\Sigma_{UC}$ of $\sim$7.3$\times$10$^{-3}$ deg$^{-2}$.
Our short exposure images of the GJ~3629~A reach $\Delta$$J$ $\sim$ 7, or $J$ $\sim$ 16.4~mag.  This is comparable
to the limiting magnitude of the 2MASS survey from which the Reid et al. compilation is based, so no correction to the
surface density is needed since the same depth is probed.
The Poisson distribution gives the probability of observing an event given an expectation value  
($\Sigma_{UC}$$\times$$\Omega$).  Since this value is small, the probability converges to the expectation value itself,
which is $\sim$6$\times$10$^{-8}$ for the 10$\farcs$2$\times$10$\farcs$2 NIRC2 narrow camera.  
GJ~3629 is part of our larger PALMS imaging survey, so the total sample size of the survey must be taken into account 
to determine if this rare event is consistent with the expected contamination rate for a given number of trials.  
When the multi-band observations of GJ~3629~AB were
made, which identified the candidate companion as a late-type object, we had imaged about 100 M dwarfs.  
Assuming the contamination rate for the survey is comparable to that of GJ 3629, the probability of a
field ultracool dwarf falling within our field of view by chance is roughly 6$\times$10$^{-6}$.  Note that this only assumes
contamination from a foreground ultracool dwarf without incorporating background late-type giants (which will increase 
the probability of contamination); the coincident proper motions of the components were also not taken into account 
(which will decrease the probability of contamination).

We also estimate the probability of contamination by a background star using the TRILEGAL population synthesis 
code (\citealt{Girardi:2005p21785}).  Here we use the default settings for the initial mass function, binary statistics, extinction values, 
and components of the galaxy.  We queried a 10 deg$^2$ region of the sky at the position of GJ~3629 in the MKO filter
system down to a depth of 26~mag in $H$ band (L. Girardi, private communication).  
For the purposes of this \textit{a posteriori} calculation, the magnitude cutoffs must reflect a range that would 
have been deemed ``interesting'' and worthy of follow-up observations.
This corresponds to substellar objects down to the detection limits of our Keck short exposure images.  Given the distance ($\sim$22~pc)
and age ($\sim$100~Myr) of GJ~3629, the \citet{Baraffe:2003p587} evolutionary models give a substellar brightness cutoff 
of $J$=11.4.  The surface density of stars with 11.4$<$$J$$<$16.4 from the models is 62 deg$^{-2}$, implying a 
probability of $\sim$5$\times$10$^{-4}$ of a background star falling in the field of view.  TRILEGAL also includes a kinematic
model that predicts stellar proper motions and has been shown to reproduce observed proper motion distributions 
(\citealt{Rossetto:2011p23783}).  If in addition to our magnitude cutoff we impose a proper motion constraint 
of $\pm$30~mas/yr in $\mu_{\alpha}$cos$\delta$ and $\mu_{\delta}$ from GJ~3629, the implied surface density is 0.01~deg$^{-2}$
and the probability of contamination is $\sim$8$\times$10$^{-8}$.  (This loose proper motion constraint allows for some orbital 
motion at smaller separations.)  Note that we have not included a color cutoff; this probability is for all background stars.
Even with the sample size of 100 targets, the probability that the candidate companion is a background object is negligibly small.
GJ~3629~B is therefore physically bound to the primary, and we can expect a third epoch of astrometry in a few years to 
show evidence of curvature in the orbital motion.

\subsection{Age}{\label{sec:age}}

 X-ray and UV emission can be used to infer the ages of solar-type stars down to stars near the 
fully-convective boundary since 
high-energy emission traces magnetic field strengths, which decay over time as a 
result of slowing rotation rates (e.g., \citealt{Skumanich:1972p23130}; \citealt{Preibisch:2005p330}).
GJ~3629~AB is detected in the 
$ROSAT$ All Sky Bright-Source Catalog (\citealt{Voges:1999p22945}) and 
both the near-UV (NUV) and far-UV (FUV) bands of the \emph{Galaxy Evolution
Explorer} ($GALEX$; \citealt{Martin:2005p23310}) space telescope as part of its 
All-sky Imaging Survey (\citealt{Morrissey:2007p22251}).

The measured count rate from $ROSAT$ is 0.127~$\pm$~0.022 sec$^{-1}$ with 
hardness ratios from the Position Sensitive Proportional Counter instrument 
of $HR1$ = --0.02~$\pm$~0.16 and $HR2$ = --0.40~$\pm$~0.20.
 This corresponds to an X-ray flux of 
1.0~$\pm$0.2$\times$10$^{-12}$~erg~s$^{-1}$~cm$^{-2}$ 
using the conversion factor from \citet{Fleming:1995p23307}, which uses the $HR1$ measurement as a 
proxy for the shape of the X-ray spectrum.  
Assuming a distance of 22~$\pm$~3~pc, GJ~3629~AB has an X-ray luminosity
of log~$L_X$ = 28.8~$\pm$~0.2~erg~s$^{-1}$.  We calculate a fractional luminosity (which
is independent of distance) of log($L_X$/$L_\mathrm{Bol}$) = --2.90~$\pm$~0.10 
using the $H$-band bolometric correction from \citet{Casagrande:2008p23483}.
Compared to the cumulative distributions of X-ray luminosities and fractional luminosities 
for young clusters from \citet{Preibisch:2005p330}, the values for GJ~3629~AB are 
comparable to cluster members with ages younger than the Hyades (625~Myr).  The large fractional X-ray
luminosity is in the saturated regime where M dwarfs cease to emit higher X-ray fluxes 
even with faster rotation rates 
(\citealt{Delfosse:1998p23131}; \citealt{Pizzolato:2003p23304}; \citealt{Wright:2011p23132}).
\citet{Shkolnik:2009p19565} derive an X-ray to $J$-band fractional luminosity of --2.19,
which is likewise higher than nearly all Hyades members for that color.  
The hardness ratios also appear to be closer to those of young moving group members and
Hyades members than the softer values of the field population (\citealt{Kastner:2003p23108}).

\begin{figure}
  \resizebox{\textwidth}{!}{\includegraphics{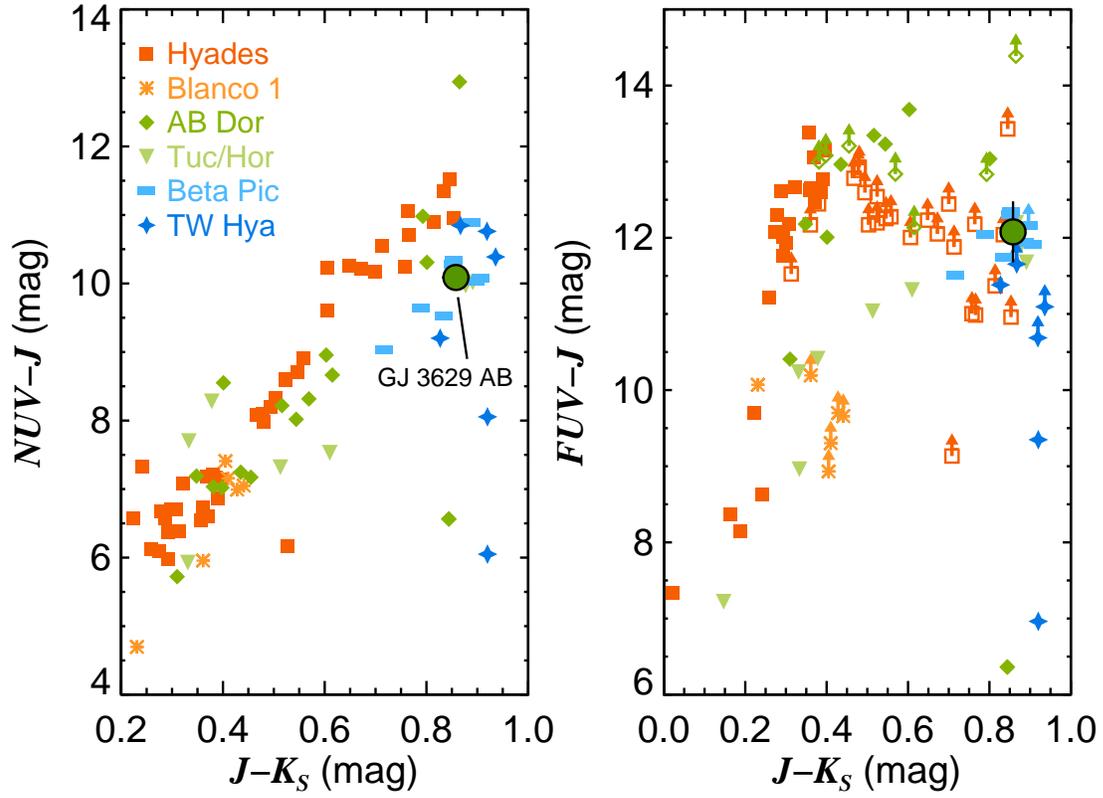}}
  \caption{$GALEX$ $NUV$--$J$ (left) and $FUV$--$J$ (right) colors as a function of 
  $J$--$K_S$ for young clusters between $\sim$8~Myr (TW Hya) and $\sim$625~Myr (Hyades).
  Young stars tend to have blue UV--NIR colors, generally falling below the Hyades sequence. 
  GJ~3629~AB (green filled circle) sits $\sim$1~mag below this envelope in $NUV$--$J$ and 
  appears to have an age confidently
  less than the Hyades.  Photometry for the cluster members are from \citet{Findeisen:2011p22756}; 
  open symbols and upward arrows indicate lower limits.  \label{fig:galex} } 
\end{figure}

Several studies have noted that young stars separate themselves from old field dwarfs
in a variety of $GALEX$-NIR/optical colors (\citealt{Findeisen:2010p20370}; \citealt{Shkolnik:2011p21923}; 
\citealt{Rodriguez:2011p21813}; \citealt{Findeisen:2011p22756}; 
\citealt{Schlieder:2012p23477}).  Figure~\ref{fig:galex} shows the location of 
Hyades ($\sim$625~Myr), Blanco~1 ($\sim$100~Myr), AB~Dor~($\sim$~100~Myr), 
Tuc/Hor ($\sim$30~Myr), $\beta$~Pic ($\sim$12~Myr), and TW~Hya ($\sim$8~Myr) members
in the $NUV$--$J$ vs. $J$--$K_S$ and $FUV$--$J$ vs. $J$--$K_S$ diagrams 
from \citet{Findeisen:2011p22756}.  The Hyades trace an upper red envelope 
with younger stars generally having bluer $NUV$--$J$ and $FUV$--$J$
colors.  The photometry for GJ~3629~AB (Table~\ref{tab:photometry}) fall $\sim$1~mag below
this sequence and have similar colors to M-type members of the $\beta$~Pic moving group
in both diagrams.  However, we note that the lack of Li observed by \citet{Shkolnik:2009p19565}
places a lower limit to the age $\sim$25~Myr based on the Li depletion models of \citet{Chabrier:1996p23529}.  
The $GALEX$ data therefore bolster a young
age for GJ~3629~AB as indicted by its X-ray emission but do not provide any tighter age constraints;
we therefore adopt the 25--300~Myr age suggested by \citet{Shkolnik:2009p19565} for this work.
The $UVW$ kinematics and $XYZ$ position of GJ~3629~AB (Table~\ref{tab:properties}) 
based on the measured radial velocity of the system (Shkolnik et al. 2012, ApJ, submitted) 
do not appear to match those of any known young moving groups.  

Since rotation rates decay over time, in principle it should be possible to infer ages from
rotation periods by empirically calibrating them to well-studied coeval clusters.
It has become clear in recent years, however, that stars are born with a large range
of rotation periods that evolve non-uniformly (e.g., \citealt{Stassun:1999p23112}).   
This is particularly true of M dwarfs with
fully convective interiors ($M$~$\lesssim$~0.3~\Msun, SpT~$\gtrsim$~M4),
whose periods increase between $\sim$10 and 200~Myr as they contract to the zero-age main sequence
and then exhibit a large spread in ages at $\gtrsim$1~Gyr (\citealt{Irwin:2011p22865}).
\citet{Hartman:2011p22788} measured a rotation period for GJ~3629~A of 3.78 days as
part of the HATNet project to find transiting extrasolar planets (\citealt{Bakos:2004p23490}).
With a mass of $\sim$0.25~\Msun \ (Section~\ref{sec:properties}), GJ~3629~A sits
near the edge of this fully convective boundary, and based on the compilation of
open cluster data from \citet{Irwin:2011p22865} the age cannot be constrained from
its rotation period alone.

Finally, we note that an upper limit for the age of the system can be estimated using 
the typical activity lifetimes for M dwarf revealed through H$\alpha$ emission.   \citet{Shkolnik:2009p19565}
detected H$\alpha$  emission (EW=--3~\AA) in GJ~3629~A, which implies an age of $\lesssim$2~Gyr
given the spectral type-activity lifetime relations derived by \citet{West:2008p19562}.  This provides
further support that the system is young.

\subsection{Spectral Type}{\label{sec:spt}}

\begin{figure}
  \resizebox{\textwidth}{!}{\includegraphics{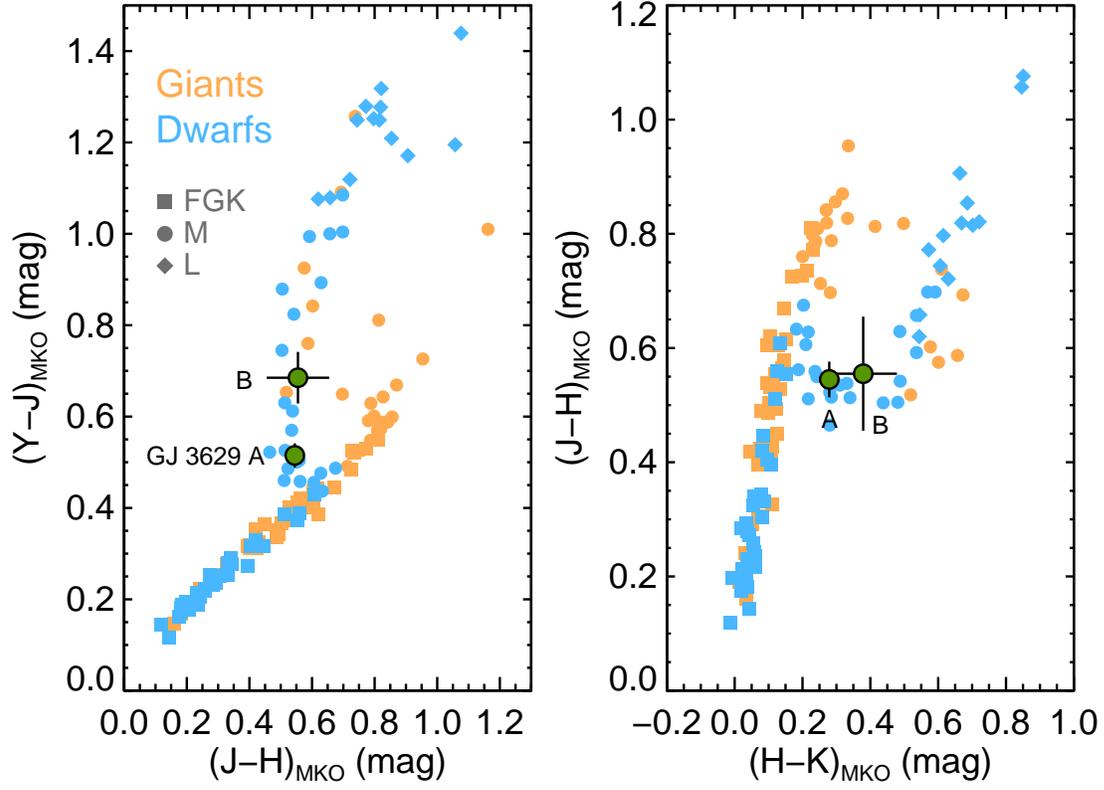}}
  \caption{Near-infrared color-color diagrams showing the positions of GJ3629~A and B 
  (solid green circles) relative to
  dwarfs (blue) and giants (orange).  GJ~3629~B is redder in $Y-J$ (left) and $H-K$ (right) than the primary and
  has colors comparable to M6--M7 dwarfs.  
  The FGK (squares) and M (circles) dwarf and giant data are 
  synthesized MKO photometry from \citet{Rayner:2009p19799} and the late-M and L dwarfs are 
  from \citet{Cushing:2005p288}. \label{fig:ccd} } 
\end{figure}

\begin{figure}
  \resizebox{\textwidth}{!}{\includegraphics{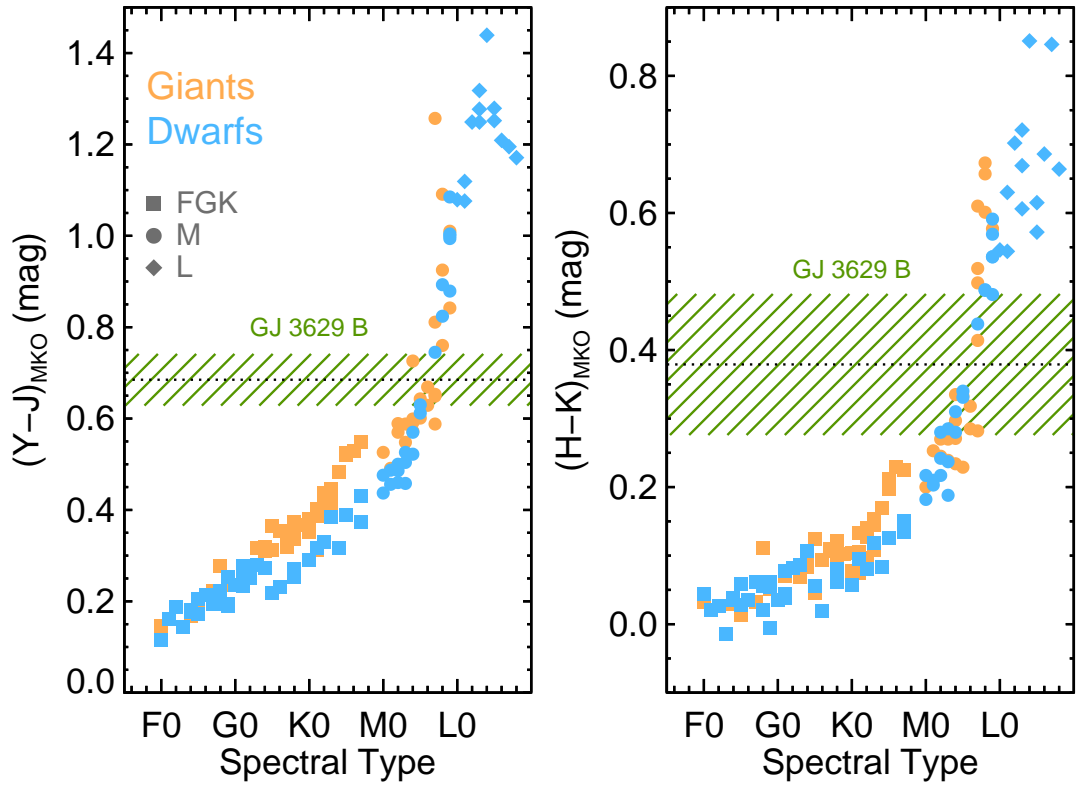}}
  \caption{Near-infrared colors for giants (orange) and dwarfs (blue) as a function of spectral type.
  The $Y-J$ (left) and $H-K$ (right) colors of GJ~3629~B suggest a spectral type of
   M7~$\pm$~2 and M6~$\pm$~4, respectively. Photometry for the dwarfs and giants are
   from \citet{Rayner:2009p19799} and  \citet{Cushing:2005p288}.   \label{fig:sptcol} } 
\end{figure}

We use the resolved photometry of GJ~3629~AB to estimate the spectral type of the companion.
Figure~\ref{fig:ccd} shows the positions of GJ~3629~A and B in $YJHK$ color-color diagrams  
relative to dwarfs (blue) and giants (orange) from \citet{Cushing:2005p288} and \citet{Rayner:2009p19799}.  
For the $Y$-band
magnitude of the primary we use the mean of the $Y$--$J$ color from the two M3V
dwarfs in \citet{Rayner:2009p19799} combined with the $J$-band magnitude of
the primary.  The $Y$-band magnitude of the companion is then computed from
our relative photometry for the system.  We estimate the intrinsic uncertainty in  $Y$--$J$ color as half the
difference of those from the two M3V objects.
 The  $J$--$H$ color of GJ~3629~B is similar to that of the primary, but the companion is redder in both
$Y$--$J$ and $H$--$K$.  Figure~\ref{fig:sptcol} shows $Y$--$J$ and $H$--$K$ colors versus spectral type
for the same comparison objects.  The $Y$--$J$ color of GJ~3629~B suggests a classification of 
M7~$\pm$~2; its $H$--$K$ color is less constraining at M6~$\pm$~4.  We therefore adopt
a spectral type of M7~$\pm$~2 for the companion.

\subsection{Physical Properties}{\label{sec:properties}}

We derive a luminosity of log($L_\mathrm{Bol}$/$L_{\odot}$)~= --1.90~$\pm$~0.13 for the 
M3.0 primary GJ~3629~A using the $H$-band bolometric correction from \citet{Casagrande:2008p23483} 
and a distance estimate of 22~$\pm$~3~pc.  
At 30, 100, and 300~Myr, the evolutionary models of 
\citet{Baraffe:1998p160} imply masses of $\sim$0.15, 0.25, and 0.30~\Msun \ and effective 
temperatures of $\sim$3120, 3370, and 3440~K, respectively.  
Because of the uncertainty in age and distance,
we assume a mass of 0.25~$\pm$~0.05~\Msun \ for the primary GJ~3629~A.

\begin{figure}
  \resizebox{\textwidth}{!}{\includegraphics{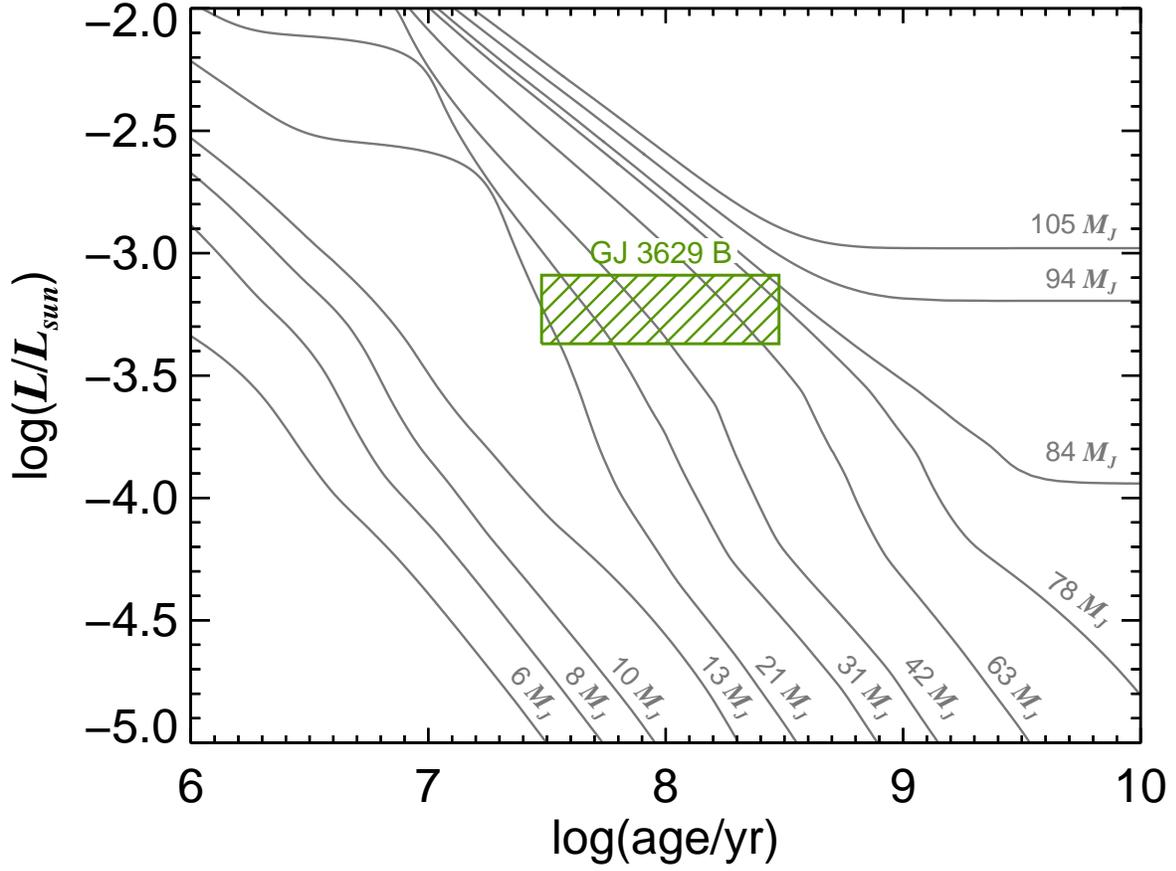}}
  \caption{Mass estimate for GJ~3629~B based on the evolutionary models of \citet{Burrows:1997p2706}.  
  The green region shows the model predictions for the age (25--300~Myr) and estimated distance 
  (22~$\pm$~3~pc) 
  of the primary, which yields  46~$\pm$~16~\Mjup \ for GJ~3629~B.   
  This is well below 
  the hydrogen burning limit ($\sim$78~\Mjup; e.g. \citealt{Burrows:2001p64}).
  \label{fig:evmods} } 
\end{figure}

We estimate a luminosity for GJ~3629~B using the $K$-band bolometric corrections
from \citet{Liu:2010p21195}, arriving at a value of  log($L_\mathrm{Bol}$/$L_{\odot}$)~=
--3.23~$\pm$~0.14 assuming a distance of 22~$\pm$~3 
and a spectral type of M7~$\pm$~2.  This translates into a mass of 46~$\pm$~16~\Mjup \ 
 based on an interpolated grid of substellar models 
from \citet{Burrows:1997p2706} as shown in Figure~\ref{fig:evmods}. Similar masses are obtained using the 
brown dwarf evolutionary models of \citet{Chabrier:2000p161}, \citet{Baraffe:2003p587}, 
and \citet{Saumon:2008p14070}.

\section{Discussion and Conclusions}{\label{sec:discussion}}
 
With an angular separation of $\approx$200~mas, GJ~3629~AB has a projected physical separation  
of 4.4~$\pm$~0.6~AU 
at a distance of 22~$\pm$~3~pc.  \citet{Dupuy:2011p22603}
calculate statistical correction factors to convert projected separations of visual binaries into 
semimajor axes using a variety of input eccentricity distributions.
Assuming their conversion factor of 1.16, which represents the case of no discovery bias and an
input eccentricity distribution from the observed population of very low-mass binaries, 
we estimate the orbital period of the GJ~3629~AB system to be 21~$\pm$~5~yr.  This raises the 
possibility of measuring a dynamical mass for the system on a modest timescale.  
Radial velocity monitoring of the system will also be feasible because 
the velocity semi-amplitude induced by the
companion is expected to be $\sim$1.0~km~s$^{-1}$ assuming an inclination of 90$^{\circ}$. 
A parallax measurement and continuing astrometric and radial velocity monitoring should 
be a high priority for this system.

A growing list of brown dwarfs now have dynamical mass measurements
(e.g., \citealt{Liu:2008p14548}; \citealt{Dupuy:2010p21117}; 
\citealt{Konopacky:2010p20823}), but very few of these also have good 
age and metallicity constraints from being companions to well-characterized stars or members
of coeval clusters.  This rare class of brown dwarfs--- representing 
both ``age benchmarks'' and ``mass benchmarks''--- functions as an excellent tool
to accurately test substellar evolutionary models (\citealt{Liu:2008p14548}).  For example, 
\citet{Dupuy:2009p15627} measured the dynamical mass of the binary brown dwarf
HD~130948~BC and used the precise age determination of the primary star (0.8~$\pm$~0.2~Gyr) to
test two of the most commonly used substellar cooling models 
(\citealt{Burrows:1997p2706}; \citealt{Chabrier:2000p161}).  They found that these models 
underpredict the luminosities of HD~130948~BC by a factor of 2--3.  
Other brown dwarfs with dynamical masses and well-constrained ``environmental''
ages (that is, not dependent on substellar cooling models) are the young ($\sim$1~Myr) 
eclipsing binary brown dwarf pair 2MASS~J05352184--0546085~AB (\citealt{Stassun:2006p23491})
and the 2.2~$\pm$~1.5~Gyr system HR~7672~B 
(\citealt{Liu:2002p18017}; \citealt{Crepp:2011p23186}).\footnote{The other systems consisting of 
one or more brown dwarfs with a dynamical mass measurement and a stellar primary all
have poor age constraints.  The triple system GJ~569 Bab  
(\citealt{Martin:2000p20363}; \citealt{Lane:2001p22845})
appears to have an age $\lesssim$1~Gyr but the estimates for the primary star vary widely 
in the literature (see \citealt{Dupuy:2010p21117} for a summary).
Likewise, various age estimates for the $\epsilon$~Indi~Bab triple system 
(\citealt{Scholz:2003p22538}; \citealt{Mccaughrean:2004p22578}; 
\citealt{Cardoso:2009p11371}) place it between $\sim$0.5--7~Gyr 
(see \citealt{Liu:2010p21195} and \citealt{King:2010p23528}).
The triple system GJ~802~AB 
(\citealt{Pravdo:2005p23764}; \citealt{Lloyd:2006p23493}) also
has a dynamical mass measurement (\citealt{Ireland:2008p19807}), but the 
age constraint based on its kinematics is rather weak at $\sim$3--10~Gyr.}  
Many of the aforementioned benchmarks used
unique methods to extract individual masses for the system 
components.  For example, \citet{Dupuy:2009p15627}
relied on the HD~130948~BC companions having nearly equal flux ratios to 
infer relative masses; \citet{Stassun:2006p23491} used the well-constrained
inclination of the eclipsing binary 2MASS~J05352184--0546085~AB to extract
individual masses from radial velocity curves; and \citet{Crepp:2011p23186} 
used radial velocity data of the primary star and visual orbit monitoring of the companion
to infer the mass of the HR~7672~B.

For GJ~3629~AB to join this rare group of benchmarks, individual masses will
have to be measured instead of a total mass for the system, which is 
what relative orbit monitoring yields.  
A stationary point source is visible $\sim$30$''$ north of GJ~3629~AB in POSS-I and POSS-II 
images from
the Digitized Sky Surveys, but it is not detected by 2MASS.  
The source (SDSS~J105120.51+360752.3) was identified by 
\citet{Schneider:2007p23494} as a quasar based on optical spectroscopy from the Sloan Digital 
Sky Survey (SDSS).  
Its optical colors from SDSS are very blue, but with a $z$-band magnitude of 18.8~mag
it should be possible to use this as a 
reference object in the near-infrared for absolute astrometry of the GJ~3629~AB system.  
Finally, we note that several observations can be made to further refine the age 
of the GJ~3629~AB  system: a parallax measurement would enable 
placement on the HR diagram and resolved near-infrared spectroscopy 
of GJ~3629~B would provide age constraints though the use of gravity-sensitive features
(e.g., \citealt{Allers:2007p66}).
Altogether, GJ~3629~AB represents a promising system for future studies.

  \acknowledgments

We thank our anonymous referee for helpful comments, Michael Cushing for providing the synthetic photometry for L dwarfs, 
Krzysztof Findeisen for the $GALEX$ photometry of young moving group members, and L\'{e}o Girardi for the TRILEGAL results.
BPB and MCL have been supported by NASA grant NNX11AC31G and NSF grant AST09-09222.
MT is supported by a Grant-in-Aid for Science Research in a Priority Area from MEXT.
It is a pleasure to thank our Keck support astronomer Hien Tran and observing assistant Jason McIlroy for their help in making this work possible.
We utilized data products from the Two Micron All Sky Survey, which is a joint project of the University of Massachusetts and the Infrared Processing and Analysis Center/California Institute of Technology, funded by the National Aeronautics and Space Administration and the National Science Foundation.
 NASA's Astrophysics Data System Bibliographic Services together with the VizieR catalogue access tool and SIMBAD database 
operated at CDS, Strasbourg, France, were invaluable resources for this work.
 The National Geographic Society-Palomar Observatory Sky Atlas (POSS-I) was made by the California 
Institute of Technology with grants from the National Geographic Society.
 The Second Palomar Observatory Sky Survey (POSS-II) was made by the California Institute of 
Technology with funds from the National Science Foundation, the National Geographic Society, 
the Sloan Foundation, the Samuel Oschin Foundation, and the Eastman Kodak Corporation.
 The Digitized Sky Surveys were produced at the 
Space Telescope Science Institute under U.S. Government grant NAG W-2166. 
The images of these surveys are based on photographic data obtained using the 
Oschin Schmidt Telescope on Palomar Mountain and the UK Schmidt Telescope. 
The plates were processed into the present compressed digital form with the permission 
of these institutions.
Finally, mahalo nui loa to the kama`\={a}ina of Hawai`i for their support of Keck and the Mauna Kea observatories.
We are grateful to conduct observations from this mountain.

\facility{{\it Facilities}: \facility{Keck:II (NIRC2)}}

\newpage


\clearpage
\newpage

\clearpage
\newpage

\clearpage
\newpage

\clearpage
\newpage

\clearpage
\newpage

\clearpage
\newpage

\clearpage
\newpage

\begin{deluxetable}{lccc}
\tablewidth{0pt}
\tablecolumns{4}
\tablecaption{Keck-II/NIRC2 Observations\label{tab:obs}}
\tablehead{
        \colhead{UT Date}   &       \colhead{}  &    \colhead{No. of}  &    \colhead{Coadds $\times$ }  \\
        \colhead{(Y/M/D)}   &       \colhead{Filter}  &    \colhead{Exposures}  &    \colhead{Exp. Time (s)}  
        }   
\startdata
2011/03/25  &    $K_S$  & 3 & 100$\times$0.11    \\
2012/03/03  &    $Y$  & 9 & 20$\times$1.0    \\
2012/03/03  &    $J$  & 12 & 20$\times$0.11    \\
2012/03/03  &    $H$  & 20 & 20$\times$0.11    \\
2012/03/03  &    $K$  & 20 & 10$\times$0.11    
\enddata
\end{deluxetable}

\begin{deluxetable}{lcccccc}
\tabletypesize{\small}
\tablewidth{0pt}
\tablecolumns{7}
\tablecaption{Keck-II/NIRC2 Astrometry of GJ~3629~AB\label{tab:astrometry}}
\tablehead{
        \colhead{Epoch (UT)}   &   \colhead{Filter}  &  \colhead{FWHM (mas)}   &    \colhead{Strehl}  &
        \colhead{Separation (mas)}    &    \colhead{PA ($^{\circ}$)}   &   \colhead{$\Delta$mag}
        }   
\startdata
2011.228   &   $K_S$  &  50.9~$\pm$~0.6  &  0.43~$\pm$~0.03  &   212.93~$\pm$~0.12  &  122.71~$\pm$~0.09   &  2.875 $\pm$ 0.012  \\
2012.171   &    $Y$      &  31.8~$\pm$~1.0  &  0.05~$\pm$~0.01  &   196.9~$\pm$~0.5      &  120.75~$\pm$~0.11    &  3.13 $\pm$ 0.04  \\
2012.171   &    $J$       & 33.9~$\pm$~1.4    &  0.12~$\pm$~0.02  &   197.8~$\pm$~0.4      &  121.2~$\pm$~0.2       &   2.96 $\pm$ 0.03  \\
2012.171   &    $H$      & 40.2~$\pm$~1.3    &  0.22~$\pm$~0.03  &   197.9~$\pm$~0.5      &  120.9~$\pm$~0.2       &  2.95 $\pm$ 0.09   \\
2012.171   &    $K$      & 49.8~$\pm$~0.4   &  0.45~$\pm$~0.02   &   199.5~$\pm$~0.6      &  120.54~$\pm$~0.12   & 2.85 $\pm$ 0.04   
\enddata
\end{deluxetable}


\begin{deluxetable}{lcc}
\tabletypesize{\small}
\tablewidth{0pt}
\tablecolumns{3}
\tablecaption{Photometry of GJ3629~AB\label{tab:photometry}}
\tablehead{
        \colhead{Property}   &    \colhead{Primary}    &    \colhead{Secondary}
        }   
\startdata
$R_\mathrm{USNO-B}$ (mag)     &     12.94\tablenotemark{a}   &  $\cdots$   \\
$I_\mathrm{USNO-B}$ (mag)     &     10.63\tablenotemark{a}    &  $\cdots$   \\
$Y_\mathrm{MKO}$ (mag)     &    9.88   $\pm$ 0.03\tablenotemark{b} &  13.01~$\pm$~0.05  \\
$J_\mathrm{MKO}$ (mag)     &     9.36 $\pm$ 0.02\tablenotemark{c} &  12.32 $\pm$ 0.04  \\
$H_\mathrm{MKO}$ (mag)     &     8.82 $\pm$ 0.02\tablenotemark{c}  & 11.77 $\pm$ 0.09 \\
$K_\mathrm{MKO}$ (mag)     &     8.54 $\pm$ 0.02\tablenotemark{c}  & 11.39 $\pm$ 0.05  \\
$K_S$ (mag)     &     8.564 $\pm$ 0.017  & 11.44~$\pm$~0.02  \\
$GALEX$ $NUV$ (mag)    &     \multicolumn{2}{c}{19.51 $\pm$ 0.11\tablenotemark{d}}    \\
$GALEX$ $FUV$ (mag)    &     \multicolumn{2}{c}{21.5 $\pm$ 0.4\tablenotemark{d}}    \\
$ROSAT$ flux (erg sec$^{-1}$ cm$^{-2}$)    &     \multicolumn{2}{c}{1.0~$\pm$0.2$\times$10$^{-12}$\tablenotemark{e}}    \\
$ROSAT$ HR1    &     \multicolumn{2}{c}{--0.02 $\pm$ 0.16\tablenotemark{e}}    \\
$ROSAT$ HR2   &     \multicolumn{2}{c}{--0.40 $\pm$ 0.20\tablenotemark{e}}    

\enddata
\tablenotetext{a}{USNO-B photometry from \citet{Monet:2003p17612}.  They estimate the photometric accuracy of the 
  catalog to be 0.3~mag.}
\tablenotetext{b}{Estimated from the $Y$-$J$ colors (0.52~$\pm$~0.01~mag) of M3 dwarfs 
    from \citet{Rayner:2009p19799}.}
  \tablenotetext{c}{Converted to the MKO system from 2MASS (\citealt{Skrutskie:2006p589}) using 
  the transformation in \citet{Leggett:2006p2674}.}
\tablenotetext{d}{$GALEX$ photometry from \citet{Morrissey:2007p22251}.}
\tablenotetext{e}{From the $ROSAT$ All-Sky Survey (\citealt{Voges:1999p22945}).  
The relation from \citet{Fleming:1995p23307} was used to convert count rate to flux.}
\end{deluxetable}

\clearpage

\begin{deluxetable}{lcc}
\tabletypesize{\small}
\tablewidth{0pt}
\tablecolumns{3}
\tablecaption{Properties of GJ~3629~AB\label{tab:properties}}
\tablehead{
        \colhead{Property}   &    \colhead{Primary}    &    \colhead{Secondary}
        }   
\startdata
Age (Myr)     &     \multicolumn{2}{c}{25--300\tablenotemark{a}}   \\
$d_{est}$ (pc)     &     \multicolumn{2}{c}{22 $\pm$ 3}   \\
Proj. Sep. (AU)     &     \multicolumn{2}{c}{4.4 $\pm$ 0.6}  \\
$\mu_{\alpha}$cos$\delta$ (mas/yr)    &     \multicolumn{2}{c}{--192.0 $\pm$ 1.0\tablenotemark{b}}   \\
$\mu_{\delta}$ (mas/yr)    &    \multicolumn{2}{c}{--48.0 $\pm$ 6.0\tablenotemark{b}}   \\
$RV$ (km/s)    &   \multicolumn{2}{c}{13.0 $\pm$~0.3\tablenotemark{c}}   \\
$U$ (km/s)    &   \multicolumn{2}{c}{--22 $\pm$ 2}   \\
$V$ (km/s)    &   \multicolumn{2}{c}{--10.9 $\pm$ 1.6}   \\
$W$ (km/s)    &   \multicolumn{2}{c}{3.1 $\pm$ 1.2}   \\
$X$ (pc)    &   \multicolumn{2}{c}{--9.9 $\pm$ 1.4}   \\
$Y$ (pc)    &   \multicolumn{2}{c}{--1.03 $\pm$ 0.14}   \\
$Z$ (pc)    &   \multicolumn{2}{c}{20 $\pm$ 3}   \\
log($L_\mathrm{X}$/$L_\mathrm{Bol}$)    &     \multicolumn{2}{c}{--2.90~$\pm$~0.1}   \\
log($L_\mathrm{Bol}$/$L_{\odot}$)    &     --1.90~$\pm$~0.13   &  --3.23~$\pm$~0.14   \\
Spectral Type     &     M3.0~$\pm$~0.5\tablenotemark{a}   &  M7~$\pm$~2\tablenotemark{d}   \\
Mass     &     0.25~$\pm$~0.5~\Msun   &  46~$\pm$~16~\Mjup   \\
$P_\mathrm{Rot}$ (days)    &     3.78\tablenotemark{e}   &  $\cdots$   \\

\enddata
\tablecomments{$UVWXYZ$ values are based on the photometric distance estimate.  $U$ and $X$ are positive toward the
Galactic center, $V$ and $Y$ are positive toward the direction of galactic rotation, and $W$ and $Z$ are positive toward the
North Galactic Pole.}
\tablenotetext{a}{\citet{Shkolnik:2009p19565}.}
\tablenotetext{b}{From the NOMAD catalog (\citealt{Zacharias:2005p23500}).}
\tablenotetext{c}{Shkolnik et al. (2012, submitted).}
\tablenotetext{d}{Estimated from $YJHK$ colors.}
\tablenotetext{e}{From the HATNet survey (\citealt{Hartman:2011p22788}.)}
\end{deluxetable}

\end{document}